\documentclass[%
 reprint,
superscriptaddress,
 amsmath,amssymb,
 aps,
]{revtex4-2}
\usepackage{mathtools}
\usepackage{aps_macros}
\usepackage{pifont}
\usepackage{graphicx}% Include figure files
\usepackage{dcolumn}% Align table columns on decimal point
\usepackage{bm}% bold math
\usepackage{hyperref}% add hypertext capabilities
\usepackage{subfigure}%
\usepackage{float}% 
\hypersetup{
     colorlinks   = true,
     citecolor    = blue
}

\hypersetup{colorlinks=true, citecolor=blue, linkcolor = magenta}

\begin{document}

\preprint{APS/123-QED}

\title{Rotating Neutron Stars: Anisotropy Model Comparison}

\author{L.~M.~Becerra}
\email{laura.becerra@umayor.cl}
\affiliation{Centro Multidisciplinario de F\'isica, Vicerrector\'ia de Investigaci\'on, Universidad Mayor,  Santiago de Chile 8580745, Chile}
 
\author{E.~A.~Becerra-Vergara}
\email{eduar.becerra@correo.uis.edu.co}
\affiliation{Grupo de Investigaci\'on en Relatividad y Gravitaci\'on, Escuela de F\'isica, Universidad Industrial de Santander A. A. 678, Bucaramanga 680002, Colombia}

\author{F.~D.~Lora-Clavijo} 
\email{fadulora@uis.edu.co}
\affiliation{Grupo de Investigaci\'on en Relatividad y Gravitaci\'on, Escuela de F\'isica, Universidad Industrial de Santander A. A. 678, Bucaramanga 680002, Colombia}

\date{\today}

%%%%%%%%%%%%%%%%%%%%%%%%%%%%%%%%%%%%%%%%%%%%%%%%%%%%%%%%%%%%%%%%%%%%%%%%%%%%%%%%%%%%%%%%%%%
\begin{abstract}

We build slowly rotating anisotropic neutron stars using the Hartle-Thorne formalism, employing three distinct anisotropy models--Horvat, Bowers-Liang, and a covariant model--to characterize the relationship between radial and tangential pressure. We analyze how anisotropy influences stellar properties such as the mass-radius relation, angular momentum, moment of inertia, and binding energy. Our findings reveal that the maximum stable mass of non-rotating stars depends strongly on the anisotropy model, with some configurations supporting up to 60\% more mass than their isotropic counterparts with the same central density. This mass increase is most pronounced in the models where the anisotropy grows toward the star's surface, as seen in the covariant model. Furthermore, slowly rotating anisotropic stars adhere to universal relations for the moment of inertia and binding energy,  regardless of the chosen anisotropy model or equation of state.

\end{abstract}
%%%%%%%%%%%%%%%%%%%%%%%%%%%%%%%%%%%%%%%%%%%%%%%%%%%%%%%%%%%%%%%%%%%%%%%%%%%%%%%%%%%%%%%%%%%

%\keywords{Suggested keywords}%Use showkeys class option if keyword
                              %display desired
\maketitle

%\tableofcontents

%%%%%%%%%%%%%%%%%%%%%%%%%%%%%%%%%%%%%%%%%%%%%%%%%%%%%%%%%%%%%%%%%%%%%%%%%%%%%%%%%%%%%%%%%%%
\section{\label{sec:intro} Introduction}
%%%%%%%%%%%%%%%%%%%%%%%%%%%%%%%%%%%%%%%%%%%%%%%%%%%%%%%%%%%%%%%%%%%%%%%%%%%%%%%%%%%%%%%%%%%

The discovery of pulsars by Jocelyn Bell in 1967 marked the first observational evidence for neutron stars (NS) \citep{1968Natur.217..709H}, revolutionizing astrophysics. Since then, progress in theory and observation has been rapid, driven by the fact that NS are the densest objects in the observable universe \citep{2000ARNPS..50..481H,glendenning2012compact}. As natural astrophysical laboratories, they offer a unique opportunity to explore matter at extreme densities, bridging the fields of nuclear physics, particle physics, and strong-field gravity.

A common assumption in NS modeling is that its internal pressure is isotropic \citep{2000ARNPS..50..481H,2016ARA&A..54..401O,2016ApJ...820...28O}. However, this assumption often breaks down due to exotic processes within the star. Strong magnetic fields \citep{2012MNRAS.427.3406F,2015MNRAS.447.3278B}, relativistic nuclear interactions \citep{1975ARA&A..13..335C}, pion condensation \citep{1972PhRvL..29..382S}, phase transitions in superfluidity \citep{sokolov1980phase}, and other exotic phases of matter \citep{1972ARA&A..10..427R,1997PhR...286...53H,2003GReGr..35.1435D} are among the primary sources of pressure anisotropy. These effects can significantly alter key observable properties of NSs, such as their mass-radius relation \citep{2021EPJC...81..698P,2021ApJ...922..149D}, maximum mass \citep{2000astro.ph.12265D,2021Ap&SS.366....9R}, moment of inertia \citep{2020EPJC...80..769R}, tidal deformability \citep{2019PhRvC.100e5804R,2019PhRvD..99j4002B}, surface redshift \citep{1974ApJ...188..657B}, and quadrupole moment \citep{2015CQGra..32n5008S,2015PhRvD..91l3008Y,2021PhRvC.104f5805R,2022PhRvD.106j3518D}. Therefore, accurately modeling pressure anisotropy, especially in rotating NS, is crucial for understanding their internal structure and observable features.

%The influence of anisotropy on stellar properties depends on the chosen model and the level of anisotropy. To study this, various anisotropic models have been developed by introducing anisotropic pressure into the stress-energy tensor, 
To study the impact of anisotropy on stellar properties, various models have been developed by incorporating anisotropic pressure into the stress-energy tensor, \citep{1982PhRvD..26.1262B,1992MNRAS.259..365B,1994GReGr..26...75G,1976A&A....53..283H,1975A&A....38...51H,1984CaJPh..62..239K,2012GReGr..44.1419M,1995AuJPh..48..635P}.  Among these, three models have gained widespread use: \textit{(i)} the \citet{1974ApJ...188..657B} (BL) model, initially introduces to analytically solve the structure equations of incompressible stars with constant density; \textit{(ii)} the \citet{2011CQGra..28b5009H} model, which defines anisotropy through a quasi-local equation of state; and \textit{(iii)} the covariant model by \citet{2019PhRvD..99j4072R}, offering a generalized framework to analyze the dynamical properties of anisotropic self-gravitating fluids and their impact on relativistic stars.

This paper systematically compares and analyzes these three anisotropic models: the Bowers-Liang model, the Horvat model, and the covariant model. We focus on their impact on key macroscopic properties of NS, including mass, radius, angular momentum, moment of inertia, and binding energy. Within the Hartle-Thorne (HT) formalism \citep{1967ApJ...150.1005H,1968ApJ...153..807H}, we solve the structural equations for slowly rotating anisotropic NS up to second order in angular velocity. Additionally, we employ three nuclear equations of state (EOS) from the literature, representing distinct NS compositions: one with nucleons only \cite{Danielewicz2009}, another with nucleons and hyperons \cite{Typel2010}, and a third including nucleons, hyperons, and quarks \cite{Kojo2022}. Our primary objective is to assess the strengths, limitations, and applicability of each anisotropic model, offering a comprehensive understanding of their implications and constraints based on observational data.

This paper is organized as follows. In Sec.~\ref{sec:RotNS}, we outline the formalism for constructing slowly rotating anisotropic stars within the framework of general relativity. Sec.~\ref{sec:AniMod} describes the characteristics, physical assumptions, and mathematical formulation of the anisotropy models. In Sec.~\ref{subsec:resA}, we analyze static configurations (zeroth order in slow rotation), focusing on mass-radius relations and maximum mass predictions for varying anisotropy strengths. Sec.~\ref{subsec:resB} provides numerical results for rotating NSs, including moment of inertia and binding energy. Finally, we summarize our findings and discuss their implications in Sec.~\ref{sec:discussion}. Throughout this work, we use geometrized units $(c = G = 1)$ and adopt the $(-, +, +, +)$ metric signature unless stated otherwise.

%%%%%%%%%%%%%%%%%%%%%%%%%%%%%%%%%%%%%%%%%%%%%%%%%%%%%%%%%%%%%%%%%%%%%%%%%%%%%%%%%%%%%%%%%%%
\section{\label{sec:RotNS} Slowly Rotating Neutron Stars: Foundations}
%%%%%%%%%%%%%%%%%%%%%%%%%%%%%%%%%%%%%%%%%%%%%%%%%%%%%%%%%%%%%%%%%%%%%%%%%%%%%%%%%%%%%%%%%%%

Modeling rotating NS is essential for understanding their structure and observable properties under extreme conditions. In \citep{Becerra:2024xff}, we extended the HT perturbative approach to derive the structural equations for slowly rotating anisotropic NSs, including terms up to second order in the angular velocity, $\Omega$ \citep[see also][]{2024PhRvD.110b4052B}. For completeness, we summarize the key assumptions of this formalism below.

We consider an anisotropic fluid in a stationary and axially symmetric spacetime. Following the HT formalism \citep{1967ApJ...150.1005H,1968ApJ...153..807H}, the spacetime geometry is described by the line element:
\begin{equation}
\begin{aligned}
\mathrm{d}s^2 =& -e^{\nu} \left[\ 1+2  \ h \ \right] \mathrm{d}\mathrm{t}^{2}   + e^{\lambda} \left[\ 1+ 2 \  \frac{m}{r}e^{\lambda} \ \right] \mathrm{d}r^2 \\ 
& + r^2 \left[\ 1+ 2 \  k \ \right] \left[\ \mathrm{d}\theta+\sin^2\theta\left(\mathrm{d}\varphi-\omega \mathrm{d}\mathrm{t}\right)^2 \ \right], \label{eq:ds2}
\end{aligned}
\end{equation}
where $\nu$ and $\lambda$ depend only on the radial coordinate $r$ and correspond to the solution of the Tolman-Oppenheimer-Volkoff (TOV) equations for a non-rotating anisotropic star (see \citep{PhysRevD.109.043025} for details). The functions $h$, $m$, $k$, and $\omega$ depend on both $r$ and $\theta$ and represent perturbative corrections due to the star's rotational deformation:
\begin{eqnarray}
h &=& h_0(r) + h_2(r) P_2(\cos\theta)+ \mathcal{O}(\Omega^4) , \\
m &=& m_0(r) + m_2(r) P_2(\cos\theta)+\mathcal{O}(\Omega^4),\\
k &=&  k_2(r) P_2(\cos\theta)+\mathcal{O}(\Omega^4), \label{eq:k}\\
\omega &=&  \omega_1(r) P'_1(\cos\theta)+\mathcal{O}(\Omega^3).
\end{eqnarray}
Here, $P_1$ and $P_2$ are the first- and second-order Legendre polynomials, respectively. The quantity $\omega = d\varphi/dt$, proportional to $\Omega$, represents the angular velocity of the local inertial frame, accounting for frame-dragging effects due to the star's rotation. The functions $h_0(r)$ and $m_0(r)$ describe the monopolar deformation, while $h_2(r)$, $m_2(r)$, and $k_2(r)$ characterize the radial dependence of the quadrupole deformation.

As previously stated, we assume the matter source is an anisotropic fluid, described by the energy-momentum tensor~\cite{misner1973,2016GReGr..48..124P,2019PhRvD.100j3006B}.
\begin{eqnarray}\label{eq:Tmunu}
T_{\alpha \beta} &=& (\epsilon + P_\perp) u_\alpha u_\beta + P_\perp g_{\alpha \beta} + (P - P_\perp) n_\alpha n_\beta, \label{T_desc}
\end{eqnarray}
Here, $P = P(r, \theta)$ is the radial pressure, $P_\perp = P_\perp(r, \theta)$ is the tangential pressure, and $\epsilon = \epsilon(r, \theta)$ is the energy density in the comoving frame of the rotating fluid. The four-velocity of the fluid, $u^{\alpha}$, satisfies the normalization condition $u^\alpha u_\alpha = -1$. The space-like vector $n^{\alpha}$ is orthogonal to $u^{\alpha}$, satisfying $n^\alpha n_\alpha = 1$ and $u^\alpha n_\alpha = 0$. For the anisotropic, axially symmetric case, the components of $u^{\alpha}$ and $n^{\alpha}$ are given by \cite{2024PhRvD.110b4052B}:
\begin{equation}\label{eq:u_vec}
\begin{aligned}
u^\alpha = & \ e^{-\nu/2}\left(1-h+ \frac{\bar{\omega}^2}{2e^{-\nu}} \ r^2\sin^2\theta \right)[1,0,0,\Omega],
\end{aligned}
\end{equation}
\begin{equation}
\begin{aligned}
n^\alpha = & \left[0 , \dfrac{r^2 e^{-\lambda/2}}{\left(r+ 2 \ e^{\lambda} \ m\right)^{1/2}}, \frac{\mathcal{Y}}{\left[\ 1+ 2 \  k \ \right]^{1/2}}, 0\right].
\end{aligned}
\end{equation}
Here, $\bar{\omega} \equiv \Omega - \omega$ is the fluid's angular velocity relative to the local inertial frame, as observed by a freely falling observer, and $\mathcal{Y} = \mathcal{Y}(r, \theta)$ is a second-order function in $\Omega$.

Rotation deforms the star and displaces the fluid. The radial pressure, energy density, and tangential pressure can be expressed up to $\mathcal{O}(\Omega^2)$ as:
\begin{eqnarray}
    P&=& P_0\left( 1 + p_{20}+p_{22}P_2(\cos\theta)\right),\\
    \epsilon &=&\epsilon_0\left( 1 + \epsilon_{20}+\epsilon_{22}P_2(\cos\theta)\right), \label{eq:energy}\\
    P_\perp &=&P_{\perp,0}\left( 1 + p_{\perp,20}+p_{\perp,22}P_2(\cos\theta)\right) \, .
\end{eqnarray}
Using the quantities defined above and applying Einstein's field equations, $G^\mu_\nu = 8\pi T^\mu_\nu$, we derive a set of first-order differential equations for the perturbation functions $m_0(r)$, $h_0(r)$, $h_2(r)$, $k_2(r)$, and $\mathcal{Y}(r, \theta)$, along with an algebraic equation for $m_2(r)$ \cite[for a detailed derivation see][]{Becerra:2024xff}. Solving these equations requires two equation of state: one relating radial pressure to energy density, $P = P(\epsilon)$, and another incorporating tangential pressure to account for the fluid's anisotropy, which will be discussed in the following section.

%%%%%%%%%%%%%%%%%%%%%%%%%%%%%%%%%%%%%%%%%
 \subsection{Equation of state}
 %%%%%%%%%%%%%%%%%%%%%%%%%%%%%%%%%%%%%%%%%%%%%%%
%

From the available EOSs in the literature, we selected SKI3 \cite{Danielewicz2009}, DD2Y \cite{Typel2010}, and QHC21 \cite{Kojo2022}, which describe matter with different compositions: nucleons; nucleons and hyperons; and nucleons, hyperons, and quarks, respectively. These EOSs support isotropic NSs with masses above $2~M_\odot$, consistent with astrophysical observations \cite{2019ApJ...887L..21R,2022ApJ...934L..17R}.  

To simplify numerical calculations, we parameterize these EOSs using the Generalized Piecewise Polytropic (GPP) method \cite{Boyle2020}. This approach divides the baryonic rest mass density range into $N$ intervals. Within each interval, from $\rho_i$ to $\rho_{i+1}$, the pressure and energy density are expressed as:
\begin{eqnarray}\label{eq:poly_eos_a}
    P(\rho) &=& K_i\rho^{\Gamma_i} + \Lambda_i \,,\\
    \epsilon(\rho) &=& \frac{K_i}{\Gamma_i-1}\rho^{\Gamma_i} +(1+a_i)\rho - \Lambda_i \,.\label{eq:poly_eos_b}
\end{eqnarray}
The fit parameters for this method are the polytropic indices, $\Gamma_i$, and the dividing densities, $\rho_i$. The constants $K_i$, $\Lambda_i$, and $a_i$ are determined by ensuring continuity in energy density, pressure, and sound speed at the dividing densities. For the high-density region (above the nuclear saturation density, $\rho_s \approx 2.4 \times 10^{14}$ g cm$^{-3}$), we use a three-zone GPP model, while for the low-density region ($\rho < \rho_s$), we employ a five-zone parameterization \cite[for details and fit parameter values, see][]{PhysRevD.109.043025}. 

\begin{figure}[t!]
    \centering
    \includegraphics[width=0.99\linewidth]{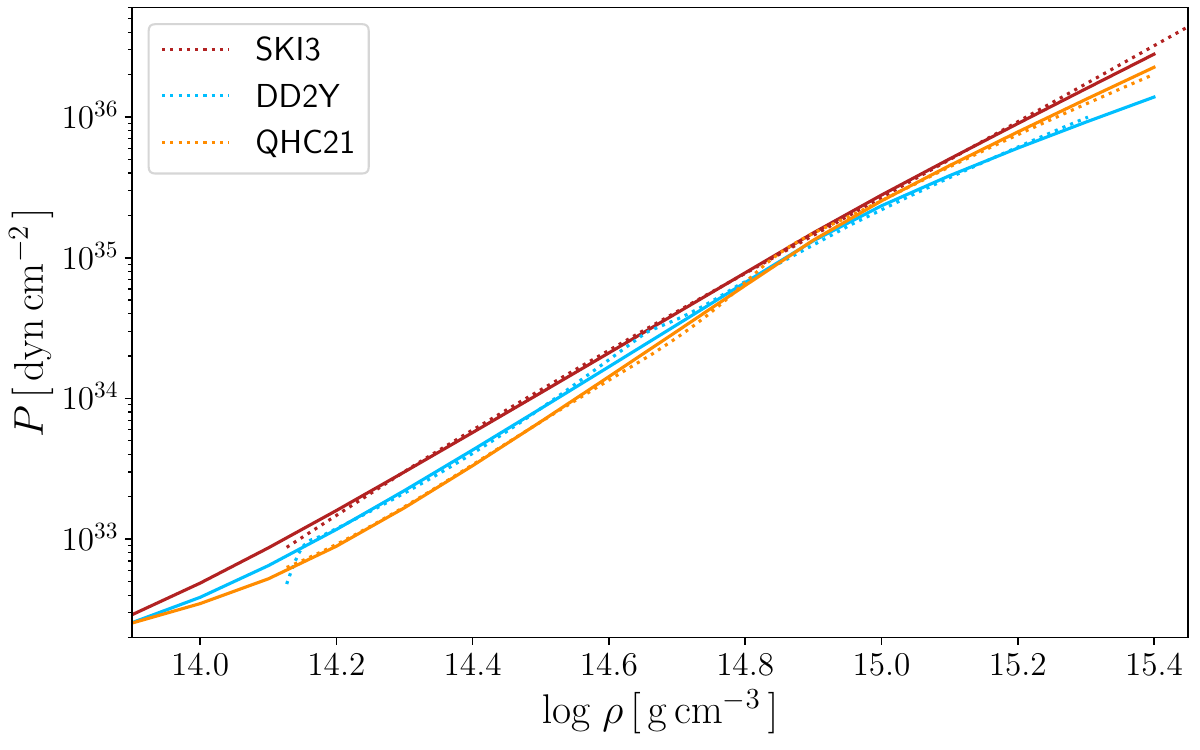}
    \caption{Pressure-mass density relation for the SKI3, DD2Y and QHC21 EOS. Dotted lines correspond to the tabulated EOS and solid lines its the corresponding GPP fit.}
    \label{fig:eos}
\end{figure}

Figure~\ref{fig:eos} compares the pressure-mass density relations of the original tabulated EOSs (dotted lines) with their GPP fits (solid lines) for the three EOSs used in this work. If we assumed a barotropic EOS (as given in equation~(\ref{eq:poly_eos_a}) and (\ref{eq:poly_eos_b})), the perturbed terms of  equation~(\ref{eq:energy}) are:
 \begin{equation}
     \epsilon_{20} = \left.\frac{P_0}{\epsilon_0}\frac{d \epsilon}{d P} \right|_{P_0}p_{20}\;\; ; \;\; \epsilon_{22}=\left.\frac{P_0}{\epsilon_0}\frac{d \epsilon}{d P} \right|_{P_0}p_{22}.
 \end{equation}

 % 

%%%%%%%%%%%%%%%%%%%%%%%%%%%%%%%%%%%%%%%%%%%%%%%%%%%%%%%%%%%%%%%%%%%%%%%%%%%%%%%%%%%%%%%%%%%
\section{\label{sec:AniMod} Anisotropy Models}
%%%%%%%%%%%%%%%%%%%%%%%%%%%%%%%%%%%%%%%%%%%%%%%%%%%%%%%%%%%%%%%%%%%%%%%%%%%%%%%%%%%%%%%%%%%

The precise relationship between energy density and radial and tangential pressures remains uncertain due to its dependence on complex microscopic factors. To address this and ensure a smooth transition between the isotropic and anisotropic regimes, many studies have introduced functional forms for the anisotropy factor: $\sigma = P - P_{\perp}$  \cite[see][and reference in there]{2023EPJC...83.1018S}. A key requirement is that the anisotropy factor must vanish at the star's center ($r\rightarrow 0$) to avoid singularities in the structure equations.  

In this study, we conduct a systematic comparison and analysis of the three leading anisotropy models documented in the literature. Our goal is to assess their strengths, limitations, and suitability for various physical scenarios, providing a comprehensive understanding of their applicability and potential constraints.

\subsection{Horvat Model}
The Horvat anisotropy model provides a significant framework for studying anisotropic fluid distributions in astrophysical systems. %It defines the anisotropy parameter as a function of radial pressure. 
A key strength of the model is its ability to naturally connect anisotropy to system compactness, making it especially useful for analyzing compact objects like NSs.  Additionally, the effect of anisotropy vanishes in the non-relativistic limit. Following \citet{2011CQGra..28b5009H}, the anisotropy factor can be expressed as
\begin{equation}\label{eq:sigma_H}
\sigma = -\lambda_H ~ \frac{2 G M }{c^2 r} P, 
\end{equation}
where $\lambda_H$ is a dimensionless parameter that controls the degree of anisotropy. In scenarios where anisotropy arises from a condensate phase of pions \cite{1972PhRvL..29..382S}, the ratio $\sigma/P$ satisfies $0 \leq \sigma/P \leq 1$, implying that the maximum pressure difference is expected to be of the order of unity. Following \cite{PhysRevD.109.043025}, we constrain the anisotropy parameter $\lambda_H$  to the range $-1 \leq \lambda_H \leq 1$.  %Given that the compactness of neutron stars (NSs) typically ranges from $0.05$ to $0.3$, the anisotropy parameter $\lambda_H$ is constrained to the range $-2 \leq \lambda_H \leq 2$ for this study \cite{2011CQGra..28b5009H, 2012PhRvD..85l4023D}. 

Finally, considering $\sigma=\sigma(P)$, for the slowly rotating configuration, the anisotropy factor can be expanded as:
\begin{equation}\label{eq:slow_sigma}
    \sigma = \sigma_0+ \sigma_{20} + \sigma_{22}P_2(\cos\theta) ,
\end{equation}
with
\begin{equation}\label{eq:slow_sigma_b}
    \sigma_{20}=\left. P_0\frac{d\sigma}{dP}\right|_{P_0} p_{20}\;\; \mathrm{and} \;\; \sigma_{22}=P_0\left. \frac{d \sigma}{dP}\right|_{P_0} p_{22}
\end{equation}

\subsection{Bowers and Liang Model}

The Bowers and Liang anisotropy model offers a foundational framework for exploring anisotropic pressure distributions in relativistic astrophysical systems. In this model, the anisotropy parameter is introduced as a function of energy density, radial pressures, and the compactness of the NS, providing a direct way to quantify deviations from isotropy. Based on the work of \citet{1974ApJ...188..657B}, the anisotropy factor is expressed as 
\begin{equation}\label{eq:sigma_BL}
\sigma = -\lambda_{BL} (\epsilon + 3P)(\epsilon + P) \left( 1-\frac{2 G M }{c^2 r} \right)^{-1} r^{2},
\end{equation}
where $\lambda_{BL}$ is a dimensionless parameter controlling the degree of anisotropy. In this model, the anisotropy is gravitationally driven and does not vanish in the non-relativistic limit. To constrain the range of possible values for $\lambda_{BL}$, we evaluate the gradient of the radial pressure and the tangential sound speed around $r\sim 0$:
\begin{equation}
   p_{0,r}\approx  \frac{2}{3}(3\lambda_{BL} - 2 \pi) (p_c + \epsilon_c) (3 p_c + \epsilon_c) r+\mathcal{O}(r^3),
\end{equation}
\begin{equation}
    c^2_{s,t}\approx c_{s,r}^2(0)\left(1-\frac{\lambda_{BL}}{\pi/3-\lambda_{BL}}+\mathcal{O}(r^2)\right),
\end{equation}
For $p_{0,r}<0$ and $0\leq c^2_{s,t}\leq 1$,  then $0\leq\lambda_{BL}\leq \pi/3$.  To implement this anisotropy in the HT formalism, we follow the approach outlined in equation~(\ref{eq:slow_sigma}) and (\ref{eq:slow_sigma_b}).

\subsection{Covariant Model}

A recent advancement in the study of anisotropic systems introduces a covariant formulation of the anisotropy parameter, offering a more geometrically and physically meaningful description. In this model, the anisotropy is expressed in a covariant manner, making it independent of coordinate choices and thus more robust for applications in general relativistic scenarios. The anisotropy parameter depends on two key physical quantities: a generic function of the energy density $f(\epsilon)$, and the projection of the radial pressure gradients along a spatial vector: $n^{\nu}$ \citet{2019PhRvD..99j4072R}. Based on this, the anisotropy parameter is given as follows:
\begin{equation}\label{eq:sigma_C}
\sigma = \lambda_{C} f(\epsilon)n^{\nu} \nabla
_{\nu} P,
\end{equation}
where $\lambda_C$ is a dimensional parameter that measures the deviation from isotropy. For the slowly rotating spacetime:
\begin{equation}
    \sigma = \sigma_0(1 + \sigma_{20} + \sigma_{22}P_2(\cos\theta) ),
\end{equation}
being $\sigma_0$ a function of order $\mathcal{O}(\Omega^0)$ and $\sigma_{20}$ and $\sigma_{22}$ functions of order $\mathcal{O}(\Omega^2)$:
\begin{eqnarray}
    \sigma_0 &=& \lambda_C f(\epsilon) \left(1-\cfrac{2M}{r}\right)^{1/2} p_{0,r}\, ,\\
    \sigma_{20} &=&\frac{m_0}{r-2M}-\frac{p_{20,r}}{p_{0,r}}\, ,\\
    \sigma_{22} &=&  \frac{m_2}{r-2M}-\frac{p_{22,r}}{p_{0,r}}\, .
\end{eqnarray}
The tangential sound speed at $r\sim 0$ is:
\begin{equation}
    c^2_{s,t}\approx c_{s,r}^2(0)\left(1-\frac{\lambda_{C}f(\epsilon_c)}{r}+\mathcal{O}(r^2)\right).
\end{equation}
Thus, to avoid singularities at the star center, we chose:
\begin{equation}\label{eq:frho}
    f(\epsilon)=\epsilon_c-\epsilon.
\end{equation}
In this case, $\lambda_C$ has dimensions of length cubed.

%%%%%%%%%%%%%%%%%%%%%%%%%%%%%%%%%%%%%%%%%%%%%%%%%%%%%%%%%%%%%%%%%%%%%%%%%%%%%%%%%%%%%%%%%%%
\section{\label{sec:res} Results}
%%%%%%%%%%%%%%%%%%%%%%%%%%%%%%%%%%%%%%%%%%%%%%%%%%%%%%%%%%%%%%%%%%%%%%%%%%%%%%%%%%%%%%%%%%%
\subsection{Non-rotating Configurations}\label{subsec:resA}

\begin{figure*}
    \centering
    \includegraphics[width=0.95\linewidth]{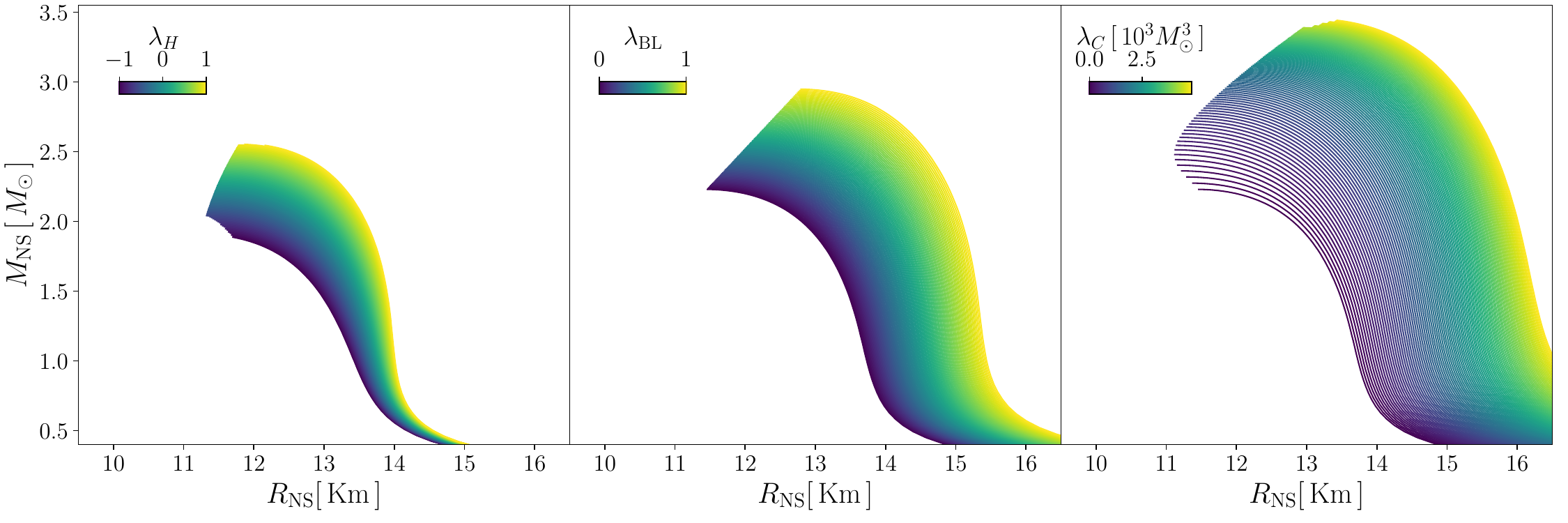}
    \includegraphics[width=0.95\linewidth]{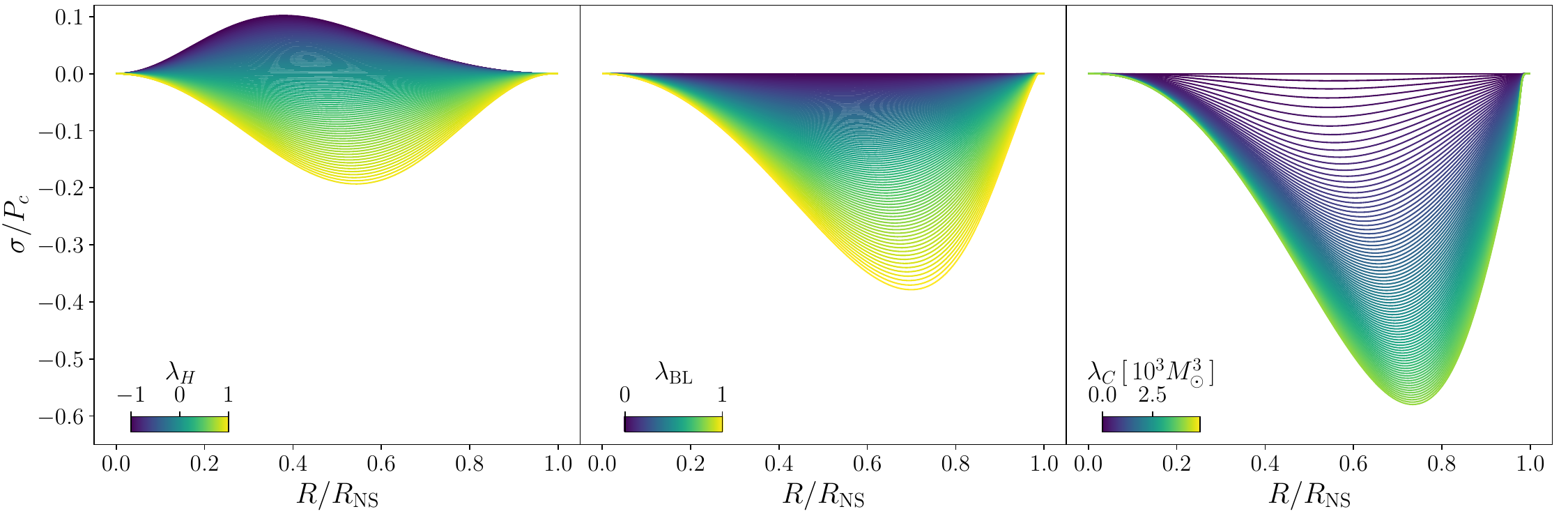}
    \caption{Upper panel: Gravitational mass as a function of radius for non-rotating  NS configurations with different anisotropy models. Lower panel: Anisotropy radial dependence for the configuration with the maximum stable mass. In both panels, the color scale corresponds to the value of the anisotropic parameter, and all configurations are computed using SKI3 EOS. }
    \label{fig:sigma}
\end{figure*}

\begin{figure*}
    \centering
    \includegraphics[width=0.98\linewidth]{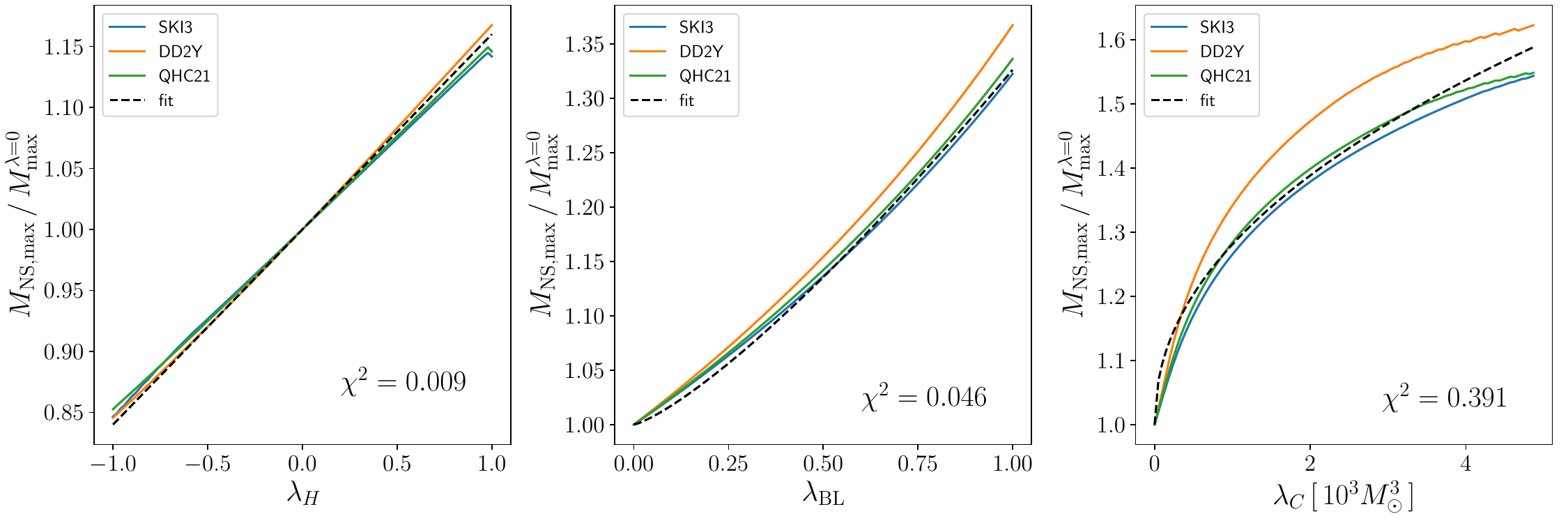}
    
    \caption{Ratio between the anisotropy maximum mass and the isotropy one as a function of the anisotropy parameter for the Horvart (left panel), Bowers-Liang (middle panel), and Covariant (right panel) anisotropy model and three NS EOS.}
    \label{fig:Mmax_static}
\end{figure*}
We construct anisotropic NS configurations in the slow-rotation approximation using the code presented in \cite{Becerra:2024xff}, an extension of the non-rotating code \cite{guzman2012revisiting,2020Ap&SS.365...43A}. This section focuses on static configurations, corresponding to the $\mathcal{O}(\Omega^0)$ order in the slow-rotation formalism.

The upper panel of Figure~\ref{fig:sigma} shows the gravitational mass as a function of the star's radius for the three anisotropy models: Horvat, Bowers-Liang, and Covariant, all computed using the SKI3 NS EOS. Only configurations satisfying the causality condition are included, ensuring that both the radial and tangential sound speeds remain below the speed of light within the star, i.e.:
\begin{equation}
    0\leq c_{s,r}^2\equiv\frac{\partial P}{\partial \epsilon} \leq 1 \quad {\rm and}\quad 0\leq  c_{s,t}^2\equiv\frac{\partial P_\perp}{\partial \epsilon}\leq 1.
\end{equation}
Along a sequence of constant anisotropy parameters in Figure~\ref{fig:sigma}, the gravitational mass of the stars increases with central energy density until it reaches a turning point, beyond which the mass begins to decrease. This turning point marks the onset of dynamical instability, leading to gravitational collapse, and defines the maximum mass of the non-rotating configuration, $M_{\rm NS,max}$ ( Figure~\ref{fig:sigma} shows only the stable configurations,  omitting the unstable branches). On the other hand, the lower panel of Figure~\ref{fig:sigma} shows the radial dependence of anisotropy inside the star for the maximum stable mass configuration, computed for each anisotropy parameter using the SKI3 EOS.

Among the three anisotropic models, only the Horvat model permits negative values of the anisotropy parameter, $\lambda_H$ ($\sigma > 0$, i.e., $P_\perp > P$), resulting in configurations less massive than their isotropic counterparts at the same central density. Conversely, all models allow positive anisotropy parameters, $\lambda > 0$ ($\sigma < 0$, i.e., $P > P_\perp$), producing more massive configurations than in the isotropic case. The covariant anisotropy model yields the most massive stars, with the largest absolute anisotropy values near the star's surface. In contrast, the Horvat model produces the least massive stars, with peak anisotropy values occurring in the star's intermediate region.

For each anisotropy model and NS EOS, Figure~\ref{fig:Mmax_static} shows the ratio of the maximum mass of an anisotropic configuration to that of the isotropic case ($\lambda = 0$) as a function of the anisotropy parameter. The figure reveals that NSs can be up to 15\% more massive with the Horvat model compared to their isotropic counterparts. This increase rises to approximately 35\% for the Bowers-Liang model and reaches 50\%–60\% for the covariant model.

The maximum mass of non-rotating anisotropic NSs can be accurately approximated by the following EOS-independent relation:
\begin{equation}\label{eq:fit_Mmax}
    M_{\rm NS,max}=M_{\rm max}^{\lambda=0}(1+a \lambda^b),
\end{equation}
where the parameters $a$ and $b$ depend on the anisotropy model and are given in Table~\ref{tab:M_max}.

\begin{table}[h!]
\centering
\begin{tabular}{ccccccccc}     
\hline
&                        & &  &                     &  &  &                   &  \\
& {\bf Anisotropy  Model}& &  & $a$                 &  &  & b                 &  \\
&                   & &  &                     &  &  &                   &  \\ \cline{2-2} \cline{5-5} \cline{8-8}
&                        & &  &                     &  &  &                   &  \\
& Horvart                & &  & $ 0.159 \pm 0.003 $ &  &  & $0.993 \pm 0.002$ &  \\
& Bowers-Liang           & &  & $ 0.326 \pm 0.005 $ &  &  & $1.267 \pm 0.039$ &  \\
& Covariant              & &  & $ 0.281 \pm 0.002 $ &  &  & $0.466 \pm 0.007$ &  \\
&                        & &  &                     &  &  &                   &  \\ \hline
    \end{tabular}
    \caption{Fit parameter for equation~(\ref{eq:fit_Mmax}). Note that in the Horvat and Bowers–Liang models, the parameter $a$ is dimensionless, whereas in the Covariant model, $a$ has dimensions  of $ (10\,  GM_{\odot}/c^2)^{-3b}$.}
    \label{tab:M_max}
\end{table}
%

%%%%%%%%%%%%%%%%%%%%%%%%%%%%%%%%%%%%%%%%%%%%
\subsection{Rotating Configurations}\label{subsec:resB}
\begin{figure*}
    \centering
    \includegraphics[width=0.95\linewidth]{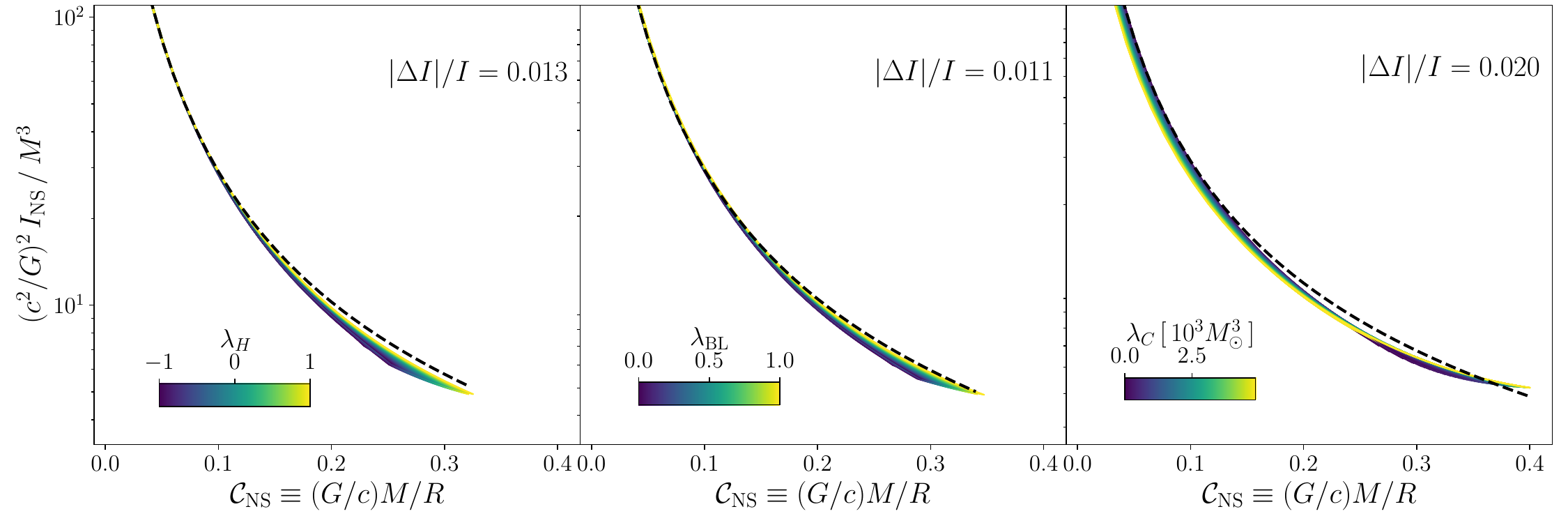}

    \includegraphics[width=0.95\linewidth]{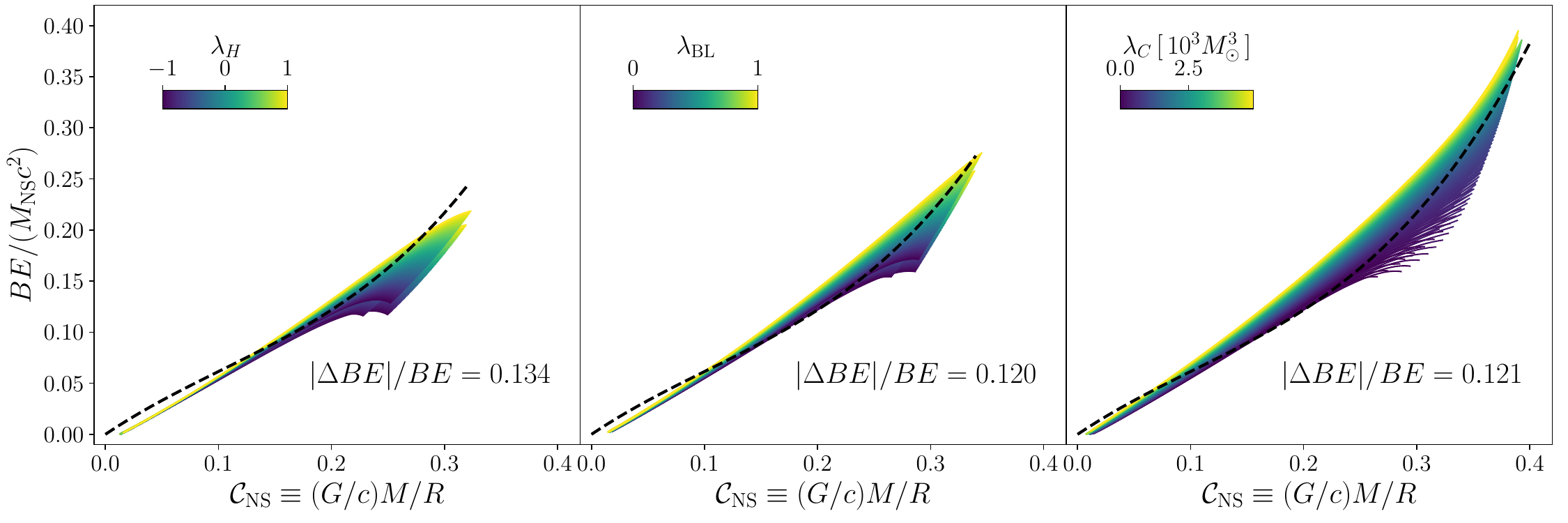}
    
    \caption{Normalized moment of inertia, $I/M^3$ (upper panel) and Binding Energy (lower panel) as a function of the non-rotating star compactness, $M/R$, for three anisotropy models and different EOS. The color scale corresponds to the value of the anisotropic parameter. The dashed black line corresponds to the fit given in equation~(\ref{eq:inertia}) for the upper panel and equation~(\ref{eq:BE}) for the lower one.}
    \label{fig:Inertia}
\end{figure*}
We now analyze the properties of slowly rotating anisotropic configurations, constructed with $\mathcal{O}(\Omega^2)$ accuracy. Our focus is on two key properties: the moment of inertia and the binding energy.

A rotating star with angular momentum $J$ and angular velocity $\Omega$ has a moment of inertia defined as:
\begin{equation}
I \equiv \frac{J}{\Omega}.
\end{equation}
Notably, up to $\mathcal{O}(\Omega^2)$, the moment of inertia is independent of the star's angular velocity.

Several studies have proposed universal relations for the NS moment of inertia \cite{1994ApJ...424..846R,2016MNRAS.459..646B}, and we have confirmed that anisotropic configurations also follow these relations \cite{Becerra:2024xff}. The upper panel of Figure~\ref{fig:Inertia} shows the normalized moment of inertia, $(c^2/G)^2 I/M^3$, as a function of the compactness of the corresponding non-rotating configuration, $\mathcal{C} = GM/c^2R$, for the three anisotropy models and all NS EOS considered in this work. The figure demonstrates that the normalized moment of inertia is nearly independent of the anisotropy model and NS EOS. Furthermore, the covariant anisotropy model produces configurations with higher compactness, reaching values close to $0.4$.

We compare our results with the relation proposed in \cite{Becerra:2024xff} for the normalized moment of inertia: 
\begin{equation}\label{eq:inertia}
\frac{I_{\rm NS}}{M^3} \approx \frac{1.019}{ \mathcal{C}} + \frac{0.225}{\mathcal{C}^{2}} - \frac{0.0038}{ \mathcal{C}^{3}} + \frac{2.3\times 10^{-5}}{\mathcal{C}^{4}}.
\end{equation}
This relation reproduces our results for different anisotropy models with an error of less than 3\%.

The binding energy of the star, defined as the energy required to assemble a stable configuration, is given by:
\begin{equation}
    BE = (M_B - M_{\rm NS}) c^2\, ,
\end{equation}
where $M_B$ is the baryonic mass and $M_{\rm NS}$ is the gravitational mass of the star. This binding energy is closely linked to the energy released via supernova neutrinos, which plays a critical role in the collapse dynamics and NS formation \cite{2001ApJ...550..426L}. In the slow-rotation approximation, up to $\mathcal{O}(\Omega^2)$, the binding energy is influenced by the monopolar deformation caused by rotation. 
 
The lower panel of Figure~\ref{fig:Inertia} shows the binding energy normalized by the gravitational mass as a function of the star's compactness for the different anisotropy models and all three NS EOS considered in this work. Our results indicate that anisotropy increases the binding energy, with the Covariant model producing configurations where the binding energy reaches up to 40\% of the total gravitational energy.

We propose a new universal fit for the binding energy, given by:
 \begin{equation}\label{eq:BE}
     \frac{BE}{M_{\rm NS}c^2}\approx  a_1\,\mathcal{C} +a_2\,\mathcal{C}^2 +a_3\, \mathcal{C}^3 .
 \end{equation}
with $a_1=0.740 \pm 0.004$, $a_2=-1.859 \pm 0.029$ and $a_3=6.000 \pm  0.056$. This relation reproduces our results with an error between 10\% and 15\%.

% The core collapse of a massive star into a NS releases more energy if the resulting NS is more massive. Future detections of supernova explosions will provide valuable observational data to constrain the properties of NS matter \cite[see][for example]{1995ApJ...445L.129B, 2017JCAP...11..036G}.

%The quadrupole moment of a rotating star, $Q$, quantifies its deviation from spherical symmetry due to rotation. In the case of an isotropic rotating star, the star becomes oblate, resulting in a negative quadrupole moment.

%Figure~\ref{fig:quadrupole} presents the normalized quadrupole moment, $\bar{Q}\equiv Q M/J^2$, as a function of the star's compactness for the three different anisotropy models.
%
%\begin{figure*}
%    \centering
%    \includegraphics[width=0.95\linewidth]{Figures/Quadrupole.pdf}
%    \caption{Normalized quadrupole moment}
%    \label{fig:quadrupole}
%\end{figure*}

%%%%%%%%%%%%%%%%%%%%%%%%%%%%%%%%%%%%%%%%%%%%%%%%%%%%%%%%%%%%%%%%%%%%%%%%%%%%%%%%%%%%%%%%%%%
\section{\label{sec:discussion} Discussions and Conclusions}
%%%%%%%%%%%%%%%%%%%%%%%%%%%%%%%%%%%%%%%%%%%%%%%%%%%%%%%%%%%%%%%%%%%%%%%%%%%%%%%%%%%%%%%%%%%

In this paper, we construct families of anisotropic NSs using three EOS and three distinct anisotropy models: the Horvat model, the Bowers-Liang model, and a covariant model. These configurations are developed within the HT formalism, which provides a perturbative framework for modeling rotating relativistic stars in General Relativity. The formalism expands the spacetime metric in powers of the star's angular velocity, $\Omega$, and solves the Einstein field equations perturbatively up to second order in $\Omega$. The zeroth-order solution corresponds to the non-rotating star, while higher-order terms incorporate rotational effects, including frame-dragging, moment of inertia, binding energy, and quadrupolar deformation.

By solving the modified TOV equations for slowly rotating configurations, we systematically examine how each anisotropy model affects key stellar properties, including the mass-radius relation, angular momentum, moment of inertia, and binding energy. 

As expected, when the anisotropy factor is positive—indicating that the tangential pressure exceeds the radial pressure—the NS configuration achieves a higher mass compared to its isotropic counterpart at the same central density.

In each model, the anisotropy strength within the star is controlled by the anisotropy parameter, primarily constrained by the radial and tangential sound speeds. Our results demonstrate that the choice of anisotropy model significantly affects the maximum stable mass of non-rotating configurations (see Equation~\ref{eq:fit_Mmax}). Notably, the covariant model (Equation~\ref{eq:sigma_C}) can produce stars up to 60\% more massive than their isotropic counterparts. This mass increase is closely associated with the anisotropy reaching its peak magnitude near the star's surface. Additionally, these effects may be further constrained by the functional form of $f(\rho)$ in Equation~(\ref{eq:sigma_C}) (see Equation \ref{eq:frho}).

Finally, we test the validity of universal relations for the moment of inertia and binding energy in slowly rotating anisotropic NS configurations with a barotropic EOS, expressed as functions of compactness (see Equations \ref{eq:inertia} and \ref{eq:BE}). These relations are independent of both the anisotropy model and EOS, with an accuracy better than 10\%.

It is worth noting that \citet{2024GReGr..56..118C} argued that anisotropy relations like those in the Horvat and Bowers-Liang models are not only \textit{ad hoc}—failing to model the underlying mechanism responsible for anisotropy—but also violate the weak equivalence principle. This principle ensures that, in the comoving frame, the EOS cannot depend on the spacetime metric. In contrast, the covariant anisotropy model is explicitly designed to uphold relativistic consistency, though it still lacks a fundamental physical justification for the origin of anisotropy.

\begin{acknowledgments}
F.D.L-C is supported by the Vicerrectoría de Investigación y Extensión - Universidad Industrial de Santander, under Grant No. 3703. E.~A.~B-V is supported by the Vicerrector\'ia de Investigaci\'on y Extensi\'on - Universidad Industrial de Santander Postdoctoral Fellowship Program No. 2025000167. 
\end{acknowledgments}

%\appendix

\bibliography{apssamp}% Produces the 

\end{document}